# An EMUSIM Technique and its Components in Cloud Computing- A Review

Dr. Rahul Malhotra [#1], Prince Jain [*2]

# Principal, Adesh institute of Technology,
Ghauran, Punjab, India

* Lecturer, Malwa Polytechnic College,
Faridkot, Punjab, India

*Abstract-* Recent efforts to design and develop Cloud technologies focus on defining novel methods, policies and mechanisms for efficiently managing Cloud infrastructures. One key challenge potential Cloud customers have before renting resources is to know how their services will behave in a set of resources and the costs involved when growing and shrinking their resource pool. Most of the studies in this area rely on simulation-based experiments, which consider simplified modeling of applications and computing environment. In order to better predict service's behavior on Cloud platforms, an integrated architecture that is based on both simulation and emulation. The proposed architecture, named EMUSIM, automatically extracts information from application behavior via emulation and then uses this information to generate the corresponding simulation model. This paper presents brief overview of the EMUSIM technique and its components. The work in this paper focuses on architecture and operation details of Automated Emulation Framework (AEF), QAppDeployer and proposes CloudSim Application for Simulation techniques.

## I. INTRODUCTION

Cloud computing has become a valuable platform for small companies to deploy their application services on a pay-as-you-go basis. To better exploit the elastic provisioning of Clouds, it is important that Cloud application developers, before deploying an application in the Cloud, understand its behavior when subject to different demand levels. This allows developers to understand application resource requirements and how variation in demand leads to variation in required resources. This information is paramount to allow proper Quality of Service (QoS) [1]. The Experimentation in a real environment is

1. Expensive as it requires a significant number of resources available for a large amount of time.
2. Time Bound as it depends on the application to be actually deployed and executed under different loads.
3. Single execution as a number of variables may affect experiment results and elimination of these influences which requires more repetition of experiments.

Due to these reasons, other techniques and process for application evaluation are preferred. To help developers to obtain more accurate models of their applications and to estimate performance and cost of the application in the Cloud, an integrated environment called EMUSIM is present. This environment introduced two such alternative techniques available to developers are EMUlation and SIMulation [1] [2]. The rest of the paper is organized as follows. Section 2 presents the EMUSIM and a brief overview of the systems. Section 3 introduces Automated Emulation Framework (AEF) and QAppDepolyer; and describes its architecture and operation details. Section 4 proposes CloudSim Application for Simulation techniques. Sections 5 differentiate both techniques based upon some major characteristics of Emulation and simulation and Section 6 concludes the paper and proposes future research directions.

## II. EMUSIM TECHNIQUE

EMUSIM combines emulation and simulation to extract information automatically from the application behavior via emulation and uses this information to generate the corresponding simulation model. Such a simulation model is then used to build a simulated scenario that is closer to the actual target production environment in application computing resources and request patterns. Information that is typically not disclosed by platform owners, such as location of virtual machines and number of virtual machines per host in a given time, is not required by EMUSIM [1][2]. EMUSIM is built on top of two software systems:

1. Automated Emulation Framework (AEF) [3] for emulation
2. CloudSim [4] for simulation

The proposed environment can also be extended to support other tools for these activities. In the rest of this section i briefly describe the technologies supporting for emulation and simulation of Cloud applications [2].

### III. AUTOMATED EMULATION FRAMEWORK (AEF)

Emulation Framework (EF) offers a convenient way to open digital files and run programs in their native computer environment. The EF is actually an automated workflow for running emulators with predefined content. Automated Emulation Framework (AEF) [3] is a framework for automated emulation of distributed systems. The target platform for AEF is a set of computer resources running a virtual machine manager [5].

The AEF [6] is an emulation testbed for distributed applications. It allows testers to describe the distributed system required by the application and automatically deploys a virtual infrastructure corresponding to a system in a local cluster. The cluster may be either homogeneous or heterogeneous regarding node's configuration, network configuration, network topology and application parameters which are stored and used by AEF. Because the tester may be unsure of the exact requirements of a platform of application, AEF allows a partial description of the environment [6][1].

In the partial description, the tester specifies, for each emulated grid site, the minimum and the maximum number of machines that are allowed to be deployed. Testers are also able to specify the amount of resources of a site in relation to another site. Another input from the tester is the limit in the resource usage by VMs and the network links accepted in the experiment. Defining the priorities in changing the number of site resources is also done through the input description file. AEF requires two XML files for an emulation experiment [6].

1. First one describes the virtual environment to be used in the experiment. It consists of one or more sites containing a number of machines connected through a virtualWAN. Therefore, description of such an environment contains characteristics of machines on each site (e.g., memory, disk, and operating system) and characteristics of the virtual WAN (latency and bandwidth).
2. Second file describes the application to be executed for each machine, which application has to be executed, and files to be transferred.

Each machine defined by the user is converted into a virtual machine that is automatically mapped onto a computer node by AEF and deployed on the chosen host. More than one virtual machine may be created in a single host, as long as the amount of resources required by the VMs does not exceed the node's capacity. Network information is used to automatically configure a virtual distributed system in the computer infrastructure, in a way that isolation among virtual sites is respected and communication between sites occurs according to WAN parameters defined in the input file. Details on each of the AEF modules, their goals, and the element in the cluster are provided in the following section.

A. *AEF Modules*

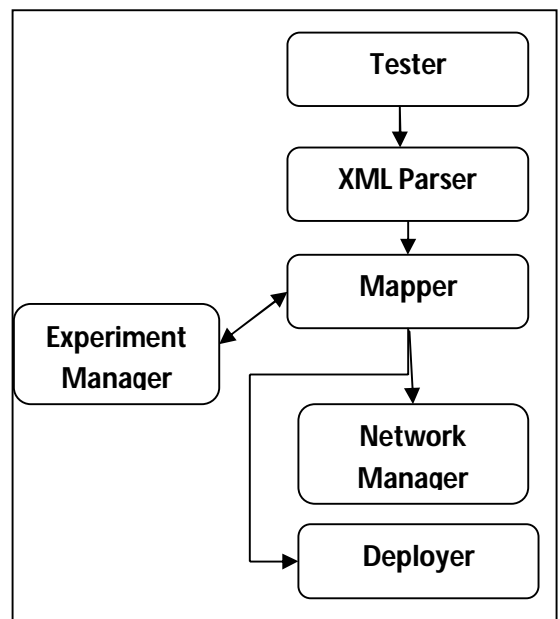

Figure 1: AEF modules

*1. XML Parser:* the major function of XML parser is to parsing of XML input files. The AEF experiment description file is known as description file. It is an XML file which contains information such as:
1. Number of virtual nodes.

2. Characteristics of virtual node such as name, VM image, memory, CPU, storage
3. Details of the connection between the nodes, latency and the bandwidth of connection.

This description file also gives information about applications to be triggered, node in which they will run, and their parameters. This file is parsed and the information is used throughout the AEF workflow to deploy nodes, configure the network and to run the experiment. The Parser module receives the description file and translates it into an internal representation of the network. It also translates the rules related to environment reconfiguration.

2. *Mapper:* the major function of mapper is to mapping of virtual environment to physical one. The Mapper uses this information and the local cluster description stored in the AEF to map both VMs in the cluster and the virtual links to physical paths in the cluster network. Virtual links are mapped to paths because VMs might be placed in hosts that are not directly connected. The virtual link is not mapped to a physical path if the VMs are mapped to the same host.
3. *Deployer:* the major function of deployer is to installation of VMs in the cluster. After the placement of VMs in the cluster is defined, VMs are actually installed by the Deployment module.
4. *Network Manager:* the major function of network manager is to configuration and managing of virtual network. The Network Manager Module configures the network, creating the virtual routes between VMs in the cluster network according to the mapping supplied by the Mapper module.
5. *Experiment Manager:* the major function of experiment manager is to installation, configuration, and monitoring of the virtual environment. When the virtual environment is built, the experiment is triggered and the environment is monitored by the Experiment Manager. To run and monitor the experiment, the Experiment Manager uses the WBEM management protocol.

IV. QAPPDEPLOYER

QoS-Aware Application Deployer [7-10] is responsible for managing the execution of applications and is started in one virtual machine by AEF during the emulation process [1].

A. *QAppDeployer Functions*
1. Focusing on parameter sweeping executions [11]
2. Responsible for mapping application tasks to VMs
3. Transferring application input data to each VM
4. Starting application execution and collecting the results from the VMs to the front-end node.
5. QAppDeployer can be configured to send a group of tasks to each executor rather than a single task.

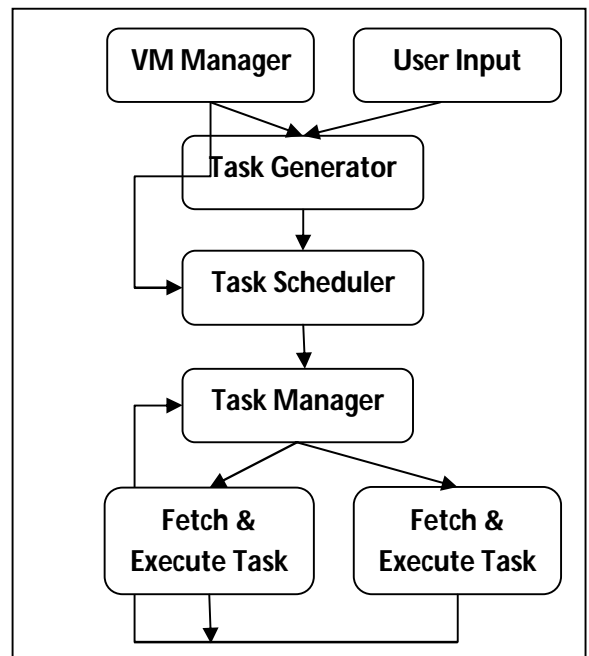

Figure 2: Modules of QAppDeployer

B. Modules of QAppDeployer
1. *VM Manager:* the AEF generates a machine file with all the available resources.
2. *Task Generator:* The task generator receives the application and its parameters.

3. *Task Scheduler:* The Task scheduler then selects a set of VMs that meet the defined requirements
4. *Task Manger:* The task manager starts an executor on each VM and transfers all required files for the application.
5. *Fetch and Execute Task:* The executors then fetch and execute tasks. Every time an executor finishes a task, it sends a message to the task manager asking for another task.

### V. CLOUDSIM

Evaluating the performance of cloud provisioning policies, services, application workload, models and resources performance models under varying system, user configurations and requirements is difficult to achieve. To overcome this challenge, CloudSim can be used. CloudSim is new, generalized and extensible simulation toolkit that enables seamless modeling, simulation and experimentation of emerging cloud computing system, infrastructures and application environments for single and internetworked clouds. In simple words, CloudSim [4] is a development toolkit for simulation of Cloud scenarios. CloudSim is not a framework as it does not provide a ready to use environment for execution of a complete scenario with a specific input. Instead, users of CloudSim have to develop the Cloud scenario it wishes to evaluate, define the required output, and provide the input parameters [1].

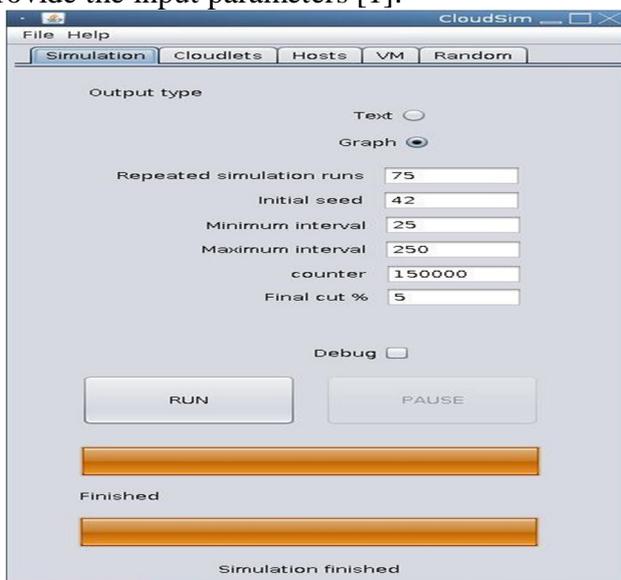

Figure 3: A snapshot of CloudSim application

CloudSim is invented and developed as "Cloudbus" Project at the University of Melbourne, Australia. The CloudSim toolkit supports system and behaviour modeling of cloud system components such as data centers, virtual machines (VMs) and resource provisioning policies. It implements generic application provisioning techniques that can be extended with ease and limited efforts. CloudSim helps the researchers and developers to focus on specific system design issues without getting concerned about the low level details related to cloud-based infrastructures and services [12]. CloudSim is an open source web application that launches preconfigured machines designed to run common open source robotic tools, robotics simulator Gazebo. Description of the scenario includes:

1. Number of data centers in the simulation.
2. Characteristics of each data center.
3. Number of hosts on each data center.
4. Hardware configuration of each host such as memory, bandwidth, number of cores, and power of each core.
5. Policies for allocation of hosts and CPU cores to virtual machines.

A. Define behavior of simulated cloud customers
1. Number of customers.
2. How and when customers request virtual machines and from which data center.
3. How customers schedule applications to create virtual machines.

B. *CloudSim Objective*

The objective of CloudSim is to provide a generalized and extensible simulation framework that enables seamless modeling, simulation and experimentation of emerging cloud computing infrastructures and application services [1].

C. *Tasks of CloudSim*

The cloud provides access to the necessary computing resources for this onetime event in a flexible manner. In general, cloud based simulation tasks can be conducted in parallel for multiple purposes:
1. validating design decisions
2. optimizing designs
3. predicting performance

4. training users
5. hosting competitions
6. improving robotics education and sharing research

While there are hourly costs associated with computing resources in a cloud environment, there is no upfront cost and little administrative effort [13]. CloudSim is used for:

1. Evaluation of resource allocation algorithms for cloud data centers.
2. Energy-efficient management of Data Centers.
3. Evaluating design and application scheduling in clouds.
4. SLA oriented management and optimization of cloud computing environments.
5. Investigation on workflow scheduling in clouds.

D. Features of CloudSim

1. Support for modeling and simulation of large scale cloud computing data centers containing hosts with different hardware capacities.
2. Support for modeling and simulation of virtualized server hosts with customizable user defined policies for host resources to virtual machines.
3. Support for modeling and simulation of energy aware computational resources.
4. Support for modeling and simulation of data center network topologies, network connections and message passing applications among simulated system elements.
5. Support for modeling and simulation of federated cloud environment that internetworks resources from both private and public domains.
6. Support for dynamic insertion of simulation elements, stop and resume of simulation environment.
7. Support for modeling of different algorithms for resource and virtual machine.
8. Support for scheduling of CPU's to VM's at virtual machine monitor level.
9. Support for a self-contained platform for modeling clouds, service brokers and provisioning.
10. Support for a virtualization engine that aids in creation and management of multiple, independent, and co-hosted virtualized services on a data center node
11. Support for flexibility to switch between space-shared and time-shared allocation of processing cores to virtualized services

These compelling features of CloudSim would speed up the development of new algorithms, methods, and protocols in Cloud computing which leads towards quicker evolution of the paradigm. With growing popularity and importance of Cloud computing, several external researchers around the world have started using CloudSim such as HP Labs, Duke University (USA), China East Jiao Tong University, National Research Center for Intelligent Computer Systems (China), and Kookmin University (Korea).

VI. *DIFFERENCE BETWEEN EMULATION AND SIMULATION TECHNIQUES*

The main difference between application emulation and simulation is the way application is represented during the evaluation process. However, the characteristics of both techniques are different to each other. Apart from this, there is some difference between them [14] [15].

Table I
Difference between Emulation and Simulation

| Emulation | Simulation |
| --- | --- |
| Emulation uses the actual software deployed in a small scale environment that models the actual production and hardware infrastructure. | Simulation allows assessment of application behaves in response to different conditions and relies on models of software and hardware for evaluation. |
| Emulation is more suitable to be used once an application software prototype is already available. | Simulation can be used in the early development stages to evaluate concepts and strategies to be used in a project. |
| Emulation requires application execution requests to be actually generated and sent to the application under test. | Simulation does not require application execution requests. |
| Emulation is little bit difficult for developers to test the model with a large number of application execution requests. | Simulation makes it easier for developers to test the model with a large number of application execution requests received from customers of the Cloud service. |
| Emulation utilization is little bit easier as it doesn't require developers to model the | Simulation utilization is difficult because it requires developers to correctly model the application |

| application behavior. | behavior If the application is not properly modeled, results obtained during simulation may not be achieved once. |
|---|---|
| Emulation typically requires more hardware resources than Simulation for the experimentation. | Simulation typically requires less hardware resources for testing which enables easier evaluation of different scenarios. |

## VII. CONCLUSION

in the Cloud computing, where access to the infrastructure incurs payments in real currency, simulation-based approaches offer significant benefits, as it allows Cloud developers to test performance of their provisioning and service delivery policies in repeatable and controllable environment free of cost and to tune the performance bottlenecks before deploying on real Clouds. This can be achieved with actual deployment, which is risky and cost ineffective. This paper, I presented EMUSIM, an architecture that combines emulation and simulation to evaluate the effect of different number of resources and patterns of requests for Cloud applications. EMUSIM was able to accurately model our two applications as a simulation model and use it to supply information about their potential performance in a Cloud provider. The AEF is an excellent alternative for the emulation of distributed systems because it uses regular and widely available clusters of workstations without requiring any special hardware to operate. It also allows a high level of control and monitoring over the executed application.